# Constrained-Realization Monte-Carlo method for Hypothesis Testing

James Theiler

*Astrophysics and Radiation Measurements Group, Los Alamos National Laboratory,*
*Los Alamos, NM 87545;* and *Santa Fe Institute, 1399 Hyde Park Road, Santa Fe, NM 87501*

Dean Prichard

*Department of Physics, University of Alaska, Fairbanks, AK 99775;* and
*Complex Systems Group, Theoretical Division, Los Alamos National Laboratory.*

**Abstract:** We compare two theoretically distinct approaches to generating artificial (or "surrogate") data for testing hypotheses about a given data set. The first and more straightforward approach is to fit a single "best" model to the original data, and then to generate surrogate data sets that are "typical realizations" of that model. The second approach concentrates not on the model but directly on the original data; it attempts to constrain the surrogate data sets so that they exactly agree with the original data for a specified set of sample statistics. Examples of these two approaches are provided for two simple cases: a test for deviations from a gaussian distribution, and a test for serial dependence in a time series. Additionally, we consider tests for nonlinearity in time series based on a Fourier transform (FT) method and on more conventional autoregressive moving-average (ARMA) fits to the data.

The comparative performance of hypothesis testing schemes based on these two approaches is found to depend on whether or not the discriminating statistic is pivotal. A statistic is "pivotal" if its distribution is the same for all processes consistent with the null hypothesis. The typical-realization method requires that the discriminating statistic satisfy this property. The constrained-realization approach, on the other hand, does not share this requirement, and can provide an accurate and powerful test without having to sacrifice flexibility in the choice of discriminating statistic.

*...provided one has access to a reasonable amount of time on a reasonably powerful computer, an exact test of significance is something one never need be without.* — G. A. Barnard, 1963

## 1 Hypothesis testing

The venerable old field of mathematical statistics provides both a language and a toolbox for dealing with the questions of inference that arise in the brave new science of chaotic dynamics. A formal framework for the "is it chaos or is it noise?" type of question which is often asked in nonlinear time series analysis can be found in the branch of statistics devoted to hypothesis testing. In this framework, one begins by asking a yes/no question about the data set of interest. For example,

- is it non-gaussian?
- is its mean significantly nonzero?
- if it is a time series, are there any temporal correlations?
- is there any nonlinear structure in the temporal correlations?
- is it chaos?

The *null hypothesis* corresponds to an answer of "no," and is the default conclusion in the lack of contrary evidence. One does not positively prove (or disprove) the null hypothesis; instead one attempts to *reject* the null hypothesis by showing that the data are unlikely to have resulted from it. This is done by computing a discriminating statistic $T$ (also called a test statistic, or a test criterion) from the data, and then inquiring whether the computed value is within a range of values that would be expected if the null hypothesis were true. If so, the null hypothesis is accepted; otherwise, it is rejected. Rejecting the null hypothesis does not prove that it is false – or even, strictly speaking, that it is improbable.[#1] What it says is that if the null hypothesis were true, then it is unlikely that data with this value of $T$ would be observed.

It is important at this point to distinguish two types of null hypotheses: *simple* and *composite*. A simple null hypothesis asserts that the given data set is a random realization of a specified unique process. The composite null specifies a family of processes, and asserts that the process that actually generated the data is a member of that family. For example, the hypothesis that data were generated by a gaussian of zero mean and unit variance is *simple*. The broader hypothesis that the data are gaussian, of unspecified mean and variance, is *composite*. To be more formal, let $\mathcal{F}$ be the space of processes under consideration, and let $\mathcal{F}_\emptyset \subset \mathcal{F}$ be the set of processes that are consistent with the null hypothesis. The null hypothesis says that the process $F$ that generated the data is an element of the set $\mathcal{F}_\emptyset$. If this set consists of a single member, then the null hypothesis is simple. Otherwise, the null hypothesis is composite, and it says that the data were generated by some process $F \in \mathcal{F}_\emptyset$ but it does not specify which $F$.

If a test rejects the null hypothesis when the null hypothesis is in fact true, this is called a "Type I error," (or a "false positive"). Usually, a test is designed with a pre-specified "size," $\alpha$, which corresponds to the expected rate of Type I errors. This parameter is adjustable; it can be made larger or smaller depending on how important it is to avoid false positives. Conventionally, $\alpha = 0.05$ is the largest value that is considered "significant". A test is "accurate" if the nominal value of $\alpha$ corresponds to the actual probability of commiting a Type I error. (Fig. 1(a) will show an example where the nominal $\alpha$ is very different from the actual rate of false positives.) Instead of choosing $\alpha$ beforehand, some authors quote the $p$-value of a hypothesis test; this is the smallest threshold $\alpha$ at which the the null hypothesis would still be rejected.

Failing to reject the null hypothesis when it is in fact false is called a "Type II error." The probability of this occuring is usually denoted $\beta$, and $1 - \beta$ is called the "power" of the test. Unlike the size $\alpha$ of the test, the power of a hypothesis test depends on how "non-null" the actual data is; that is, it depends on the actual underlying true process. Thus, one speaks

---

[#1] This is in contrast to the Bayesian approach, which *does* assign probabilities to hypotheses. Here, one begins with a *prior* probability $P(H)$ that the hypothesis $H$ is true, computes a conditional probability $P(D|H)$ that the data $D$ would be observed given that $H$ is true, and applies Bayes theorem to write $P(H|D) \propto P(D|H)P(H)$ as the *posterior* probability that $H$ is true, given an observation of data $D$.

of a test having power against a certain "alternative."

## 1.1 Monte-Carlo methods

In its classical incarnation, hypothesis testing involves choosing a discriminating statistic $T$ which is carefully tailored to match the null hypothesis. For a given null hypothesis, one chooses a $T$ for which it is straightforward (either by calculation or by table lookup) to obtain the range of values of $T$ associated with 95% ($\alpha = 0.05$) of the distribution. For this reason, there has traditionally been a preference for statistics $T$ with standard (or "standardized") distributions. However, the need for this restriction has to some extent been superseded by computer-intensive methods, in which the distribution of $T$, and its 95% confidence range, can be accurately estimated by direct Monte-Carlo simulation. This seems to have first been suggested by Barnard [1], in a brief paragraph discussing another paper, and then more fully developed by Hope [2] and others [3–7]. The idea is to compute values of $T$ for many different realizations of the null, and to empirically estimate the distribution of $T$ from this ensemble of values. Monte-Carlo methods have become even more popular for the related problem of estimating confidence intervals [8–12]; see Efron's 1979 manifesto [8] for an early and forceful argument in favor of replacing cumbersome derivations and narrow assumptions with straightforward (and increasingly cheap) computation.

We need to be careful to distinguish the quite different problems of estimating confidence intervals and testing null hypotheses. The problems are certainly related, and computationally-intensive Monte-Carlo and resampling techniques have proved valuable for both of them. As Fisher and Hall [6] note, "there are close links between bootstrap methods for testing and for interval estimation, [but] there are important, explicit differences which call for a specialized treatment of the bootstrap testing problem." In the first case, a statistic of some intrinsic interest (*e.g.*, the mean or mutual information or fractal dimension) is computed for the data, and the goal is to find "error bars" on the computed value which enclose (with some probability) the actual mean of the true underlying distribution. In the second case, there is a specific, carefully stated null hypothesis, and the goal is to test whether the data are consistent with that hypothesis.

## 1.2 The importance of being pivotal

While Monte-Carlo simulations effectively solve the problem that a distribution $T$ may not have a simple closed form solution, they are not enough, by themselves, to solve the full hypothesis testing problem. The difficulty arises when the null hypothesis is *composite*, and it is not immediately clear which process (in the family of processes covered by the null hypothesis) one should be simulating.[#2]

For this reason, it is usually demanded that $T$ be "pivotal," which means that the distribution of $T$ is the same for all members $F$ of the family $\mathcal{F}_\emptyset$ of processes consistent with the null hypothesis. In practice, this often turns out to be a very stringent criterion, and in many cases is satisfied only in the asymptotic limit as the size $n$ of the data set approaches infinity. But if $T$ is pivotal, then it doesn't matter which $F \in \mathcal{F}_\emptyset$ is used as the basis for generating realizations; all serve equally well, and it is valid to compare the obtained distribution of $T$ to the value of $T$ obtained for the data set of interest.

---

[#2] In fact, as we will see later, the problem is more difficult than that. One cannot in general test a composite null hypothesis with simulations which are typical realizations any single process. Furthermore, if we try to define a distribution of processes, and then simulate typical realizations of typical elements of the distribution, that will only make things worse.

The importance of using a pivotal statistic has been strongly emphasized by a number of authors. Hall and co-workers [4, 6, 13] showed that a Monte-Carlo test could be much more accurate than a corresponding asymptotic test *when the discriminating statistic was pivotal.* Beran [14] makes the same point, and provides a bootstrap approach for refining a statistic to make it more nearly pivotal. This is called "prepivoting" and can be very computationally intensive.

*1.3 The importance of being* non-*pivotal*

In recent years, there has been an increased interest in testing sometimes relatively uncomplicated null hypotheses, such as that the data arise from a linear stochastic process, in situations where alternatives may be relatively exotic, such as low-dimensional chaos. In this situation, there are two reasons one may want to use a complicated discriminating statistic. One reason is that a complicated test may be more powerful against a complicated alternative. For example, deterministic chaos can be distinguished from a stochastic process by its predictability; a good discriminating statistic in this case may be the error in a nonlinear prediction algorithm, and these algorithms can be very sophisticated. A second reason is that the complicated discriminating statistic may correspond to a physically interesting variable, like fractal dimension[15] or Lyapunov exponent[16]. In this case, one is able not only to *formally* test the null hypothesis, but at the same time to *informally* check whether the estimate of fractal dimension or Lyapunov exponent is corrupted by artifacts that can be explained within the simpler framework of the null hypothesis, *e.g.,* by linear correlations[#3] in the data.

The purpose of this paper is to discuss an approach to hypothesis testing that is based on a Monte-Carlo scheme in which the surrogate data sets are not "typical realizations" of a specific process, but instead are what we call "constrained realizations." The method behaves as if the discriminating statistic were pivotal, even in those cases where it is not. Since the statistic need not be pivotal, the data analyst has more flexibility in designing a test that may be powerful against even relatively exotic alternatives. We should be careful to note that we are advocating the constrained-realization approach only for hypothesis testing; it in general would not be appropriate for estimating confidence intervals.

In the next sections, we will illustrate the idea with two simple examples, corresponding to the first and third questions in our original list. We will discuss the fourth question, on testing for nonlinearity in data, at greater length in Section 3. Though the question at the bottom of our list – is it chaos? – is what motivates our interest in nonlinearity, this fifth question is too difficult for us to address directly.

## 2 Two "trivial" examples

The easiest way to explain the difference between these two approaches is by example. We will consider two cases: the first is a test for nongaussian data in a distribution of independent data points, and the second is a test for serial dependence in a time series. We should emphasize that our purpose here is not to propose new and improved statistical tests for these two situations, but merely to use these easily understood examples to illustrate the difference between "typical" and "constrained" realizations.

---

[#3]One such artifact that affects estimates of correlation dimension is described in [17]; the effects of autocorrelation on estimates of Lyapunov exponent are discussed by Daemmig and Mitschke [18, 19].

## 2.1 Are the data gaussian?

Let us take as our null hypothesis that data are gaussian. For the purposes of this example, we will implicitly assume that the data arise from an independent and identically distributed (IID) process, though in the the next section we will consider ways of testing IID as a null hypothesis in its own right. In general, the variables that parameterize the class of processes satisfying the null hypothesis are called the "nuisance parameters," and in this case the nuisance parameters are the mean $\mu$ and the variance $\sigma^2$.

### 2.1.1 Simple null hypothesis

Suppose we know $\mu = \mu_o$ and $\sigma = \sigma_o$ beforehand. The null hypothesis, then, is that the data arise from a single pre-specified process; in this case, independent sampling of the normal $N(\mu_o, \sigma_o^2)$ distribution.

To test this null hypothesis, choose a discriminating statistic $T$. A common choice is the sample average of some function of the individual data points; e.g.,

$$T = \overline{x^4} = \frac{1}{n} \sum_{i=1}^{n} x(i)^4 \qquad (1)$$

but more general statistics are also possible[#4], such as

$$T = \text{fraction of pairs } x(i), x(j) \text{ such that } |x(i) - x(j)| < r \qquad (2)$$

and so on. In general, $T$ is just a scalar function of the $n$ arguments $X = (x(1), \ldots, x(n))$.

Having chosen the discriminating statistic $T$, compute the value $t_o$ of the statistic for the data set in question. Next generate a number $B$ of surrogate data sets ($X_k, k = 1, \ldots, B$), which are just random realizations of the null process $N(\mu_o, \sigma_o^2)$. Compute the value $t_k$ of the statistic for each of the $k = 1, \ldots, B$ surrogate data sets. Finally, check to see whether $t_o$ is on the tail of the empirical distribution of $T$ given by the surrogate data. In particular, for a two-sided test, reject the null hypothesis at the level $\alpha$, if $t_o$ is observed among the largest $(B+1)\alpha/2$ or the smallest $(B+1)\alpha/2$ in the sorted list that includes $t_o$ as well as $t_1, \ldots t_B$. By construction, the probability of rejecting the null if the null is true (that is, the "size" of the test) is given by $\alpha$; this is — as Barnard promised in 1963 — an exact test for significance.

Note that $B + 1$ must be at least as large as $2/\alpha$, and is usually taken to be an integer multiple of $2/\alpha$. (For a one sided test, it only has to be a multiple of $1/\alpha$.) Apart from these caveats, the size of the test is independent of $B$; however, the power of the test improves (slightly) with increasing $B$ for reasons that are very well explained in Refs. [2, 20]. In all the numerical examples in this paper, we take $B = 39$ surrogates and aim for $\alpha = 0.05$ by rejecting the null when the original data has a discriminating statistic which is either larger than all the surrogates, or smaller than all the surrogates.

This example is straightforward, but arguably unrealistic because the null hypothesis is a single stochastic process. Usually, we do not know $\mu$ and $\sigma$ beforehand, and the null hypothesis is more generally stated: Are the data consistent with a process $N(\mu, \sigma^2)$ for

---

[#4]One problem with simple statistics such as the fourth moment in Eq. (1) is that *some non*gaussian distributions have exactly the same fourth moment as a gaussian does. The test will fail to distinguish these distributions from gaussian; in other words, it will lack power against those distributions. For this reason, some people prefer tests based on the empirical distribution function, such as the Kolmogorov-Smirnov statistic, which have some power against all nongaussian distributions, at least for large enough $n$.

*some* $\mu$ and $\sigma$; or: Can we reject $N(\mu, \sigma^2)$ for *all* $\mu$ and $\sigma$? This is a more subtle question than whether or not to reject $N(\mu_o, \sigma_o^2)$ for a *given* $\mu_o$ and $\sigma_o$. Testing against these more general (composite) null hypotheses requires more care. Below, we will describe how this can be done for the gaussian null by using surrogate data sets which are "typical realizations" of a given stochastic process, and how this is done with "constrained-realization" surrogate data.

### 2.1.2 Typical realizations

We use the term "typical realizations" to refer to the following Monte-Carlo method of testing a null hypothesis. First, the nuisance parameters $\mu$ and $\sigma$ are estimated from the original data $X_o$; call $\hat{\mu}_o$ and $\hat{\sigma}_o$ the estimated values. Then, we test the *simple* hypothesis that the data were generated by $N(\hat{\mu}_o, \hat{\sigma}_o^2)$. That is, we generate surrogate data with random realizations of $N(\hat{\mu}_o, \hat{\sigma}_o^2)$, and see whether $t_o$ is on the tail of the distribution of values obtained for all the surrogates.

This approach is known to be problematic if the discriminating statistic $T$ depends on the variables $\mu$ and $\sigma$ that parameterize the null hypothesis. Therefore, one usually demands that $T$ be a pivotal statistic (or at least that it be pivotal in the large $n$ limit), by which it is meant that the distribution of $T$ under the null hypothesis does not depend on $\mu$ or $\sigma$.

For example, consider two discriminating statistics

$$T = \overline{x^4} \qquad (3)$$
$$T' = \overline{(x-\overline{x})^4}/\overline{(x-\overline{x})^2}^2 \qquad (4)$$

where we write $\overline{f(x)}$ to denote the sample mean $(1/n)\sum_{i=1}^n f(x(i))$.[#5] The second of these is invariant to changes in translation or scale; therefore it has the same distribution for all gaussians, regardless of $\mu$ and $\sigma$. Thus it is pivotal.

In Fig. 1(a,b), we demonstrate the importance of using pivotal statistics in this typical-realizations context. We perform a numerical experiment comparing the accuracy and power of tests of a gaussian null based on the discriminating statistics given by Eq. (3) and Eq. (4). Fig. 1(a) shows that the nonpivotal statistic (Eq. (3), solid line in figure) has much smaller "size" than its nominal value of $\alpha = 0.05$. This is arguably good because it implies that there is a very low probability of commiting a Type I error, but it can also be construed as an inaccuracy since the probability of a Type I error is an adjustable parameter that one should be able to specify beforehand. In contrast, for the pivotal statistic (Eq. (4), dotted line in figure) the size is equal to the nominal value $\alpha = 0.05$. Fig. 1(b) shows what happens when the data arises from a uniform (*i.e.*, non-gaussian) distribution; we see that the pivotal statistic (dotted line) is far more likely than the nonpivotal statistic (solid line) to correctly reject the null hypothesis. The error bars in Fig. 1 were estimated using $\sqrt{p(1-p)/N}$ where $N$ is the number of trials [21].

### 2.1.3 Constrained realizations

The constrained-realization surrogate data method takes a slightly different tack. As before, we will use the data to estimate the nuisance parameters, and again, we will call these estimates $\hat{\mu}_o$ and $\hat{\sigma}_o$. However, instead of generating typical realizations of the $N(\hat{\mu}_o, \hat{\sigma}_o^2)$, we will restrict our interest to those realizations whose estimators for $\mu$ and $\sigma$ exactly match

---

[#5] In our numerics, we really used the statistic $(n-1)T'/n$, since we used $\hat{\sigma}^2 = (1/(n-1))\sum_{i=1}^n (x(i)-\overline{x})^2 = (n/(n-1))\overline{(x-\overline{x})^2}$ in the denominator, but this distinction will have no effect on the size/power curves.

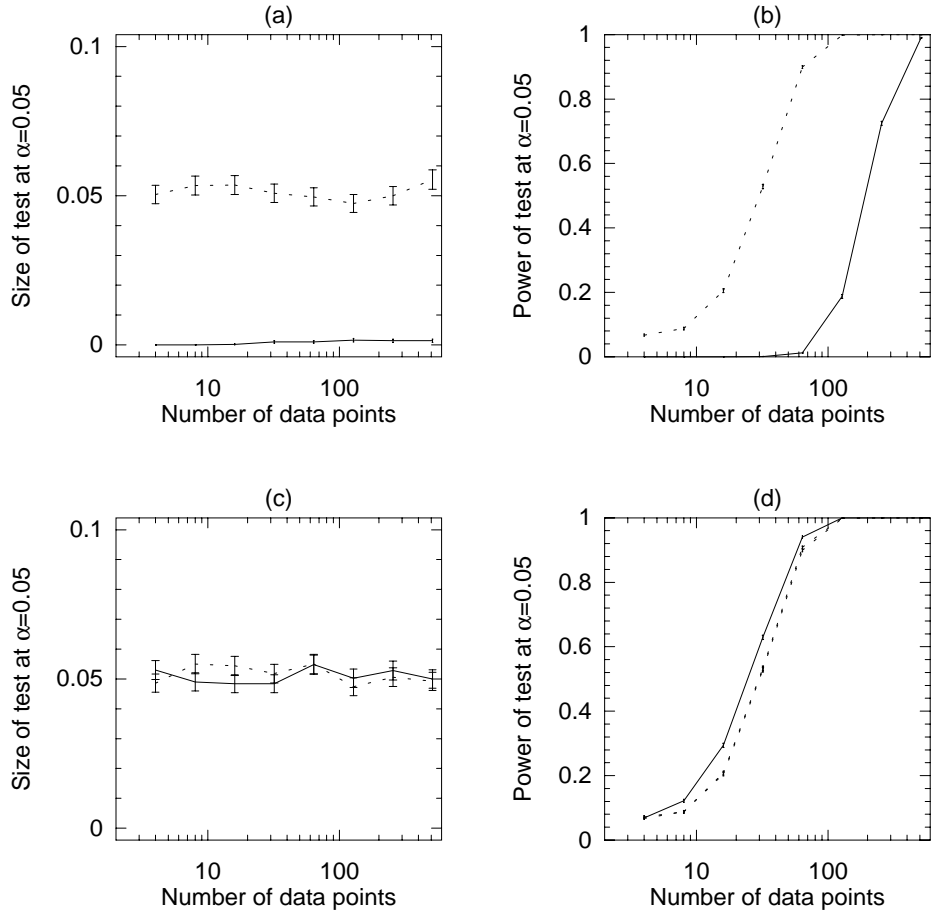

**Fig. 1.** Numerical studies of the performance of various tests of the null hypothesis that the data are gaussian: size and power are plotted against the number $n$ of points in the data set. The solid line is for the nonpivotal statistic in Eq. (3), and the dotted line is for the pivotal statistic in Eq. (4). Panels (**a,b**) are for the "typical realization" approach, and panels (**c,d**) are for the "constrained realization" approach. Plotted is the fraction of trials (out of 5000) in which the null hypothesis was rejected at the $\alpha = 0.05$ level. For the power curves, the test data came from a uniform distribution $[-\sqrt{3}, \sqrt{3}]$, and for the size curves the data came from a normal $N(0, 1)$ distribution. In the typical-realization scenario, the "size" of the pivotal statistic is almost exactly equal to the nominal value $\alpha = 0.05$, while the size of the nonpivotal statistic is practically zero. In looking at panel (**b**), we see clearly that the pivotal statistic is more powerfully able to discriminate between the gaussian null and the non-gaussian alternative. For the constrained-realization surrogates, both (**c**) the size and (**d**) the power of the nonpivotal statistic are about the same as the pivotal statistic. Note that the pivotal statistic has the same power/size properties in the constrained-realization test as it does in the typical-realization test.

those of the original data. In other words, we require $\hat{\mu}_s = \hat{\mu}_o$ and $\hat{\sigma}_s = \hat{\sigma}_o$, where $\hat{\mu}_s$ and $\hat{\sigma}_s$ are the estimates of $\mu$ and $\sigma$ obtained for the surrogate data set. How to obtain such a restricted set of surrogates for more general distributions and parameters is not a trivial issue, and we refer the reader to a preprint by Davison *et al.*[22] for further discussion.[#6]

In the case of gaussian data, however, it is easy to generate constrained-realization data sets. First, generate a typical realization of any gaussian process (we use $N(\hat{\mu}_o, \hat{\sigma}_o^2)$, but even $N(0,1)$ would work), and then rescale the data. So, if $X_k$ is a typical realization of a gaussian process with sample mean $\hat{\mu}_k$ and sample variance $\hat{\sigma}_k^2$, then rescale the data according to

$$x'(i) = \hat{\mu}_o + (x(i) - \hat{\mu}_k)\hat{\sigma}_o/\hat{\sigma}_k. \tag{5}$$

Then $X_k'$ is the constrained-realization surrogate.

For these surrogate data sets, as Fig. 1(c) and Fig. 1(d) show, the probability of rejecting the null hypothesis is about the same for the pivotal (dashed line) and the nonpivotal (solid line) statistic. For gaussian distributions (when the null is true), both statistics reject the null at the nominal rate of $\alpha = 0.05$. For the datasets generated by a nongaussian (in this case, uniform) distribution, both statistics have approximately the same power.

Oddly enough, though, the two statistics do have slightly (but significantly) different power against the uniform distribution. To understand this a little better, recall that the constrained realization method used for this example consisted of first making a typical realization $X_k$ of $N(\hat{\mu}_o, \hat{\sigma}_o^2)$, and then rescaling according to Eq. (5) to obtain the constrained realization $X_k'$.

Now computing the non-pivotal statistic $T = \overline{x^4}$ (from Eq. (3)), on a constrained realization $X_k'$ is equivalent to computing:

$$T'' = \overline{(x - \hat{\mu}_k)^4}\hat{\sigma}_o^4/\hat{\sigma}_k^4 + 4\overline{(x - \hat{\mu}_k)^3}\hat{\mu}_o\hat{\sigma}_o^3/\hat{\sigma}_k^3 + 6\frac{n-1}{n}\hat{\mu}_o^2\hat{\sigma}_o^2 + \hat{\mu}_o^4 \tag{6}$$

on the original typical realization $X_k$. We can also write this in the form:

$$T'' = c_1 T' + c_2 \overline{(x - \hat{\mu}_k)^3}/\hat{\sigma}_k^3 + c_3 \tag{7}$$

where $c_1$, $c_2$, and $c_3$ are constants, and $T'$ is the pivotal statistic defined in Eq. (4). It is clear that $T''$ is unchanged by scale or translation, and therefore it too is pivotal. So in this case, using constrained realizations with a nonpivotal statistic is equivalent to using typical realizations with a pivotal statistic.

Of course, the pivotal statistic in Eq. (7) is not the same one as the pivotal statistic in Eq. (4). It is perhaps not too surprising, then, that the two statistics do not have exactly the same power against the particular nongaussian distribution we chose for our numerical experiment.

## 2.2 Testing for IID: Resampling with and without replacement

In the previous section, the null hypothesis had exactly two nuisance parameters; here we will consider a null hypothesis in which the distribution itself is the nuisance "parameter." Given a time series $X_o = (x_o(1), \ldots, x_o(n))$, we ask whether there are any temporal correlations at all in the data by taking as our null hypothesis that the data are independent and

---

[#6] Let us remark that another paper by Davison *et al.*[23] introduced a "balanced bootstrap" which has some of the same flavor as the constrained realization, but it is more appropriate for estimating confidence intervals.

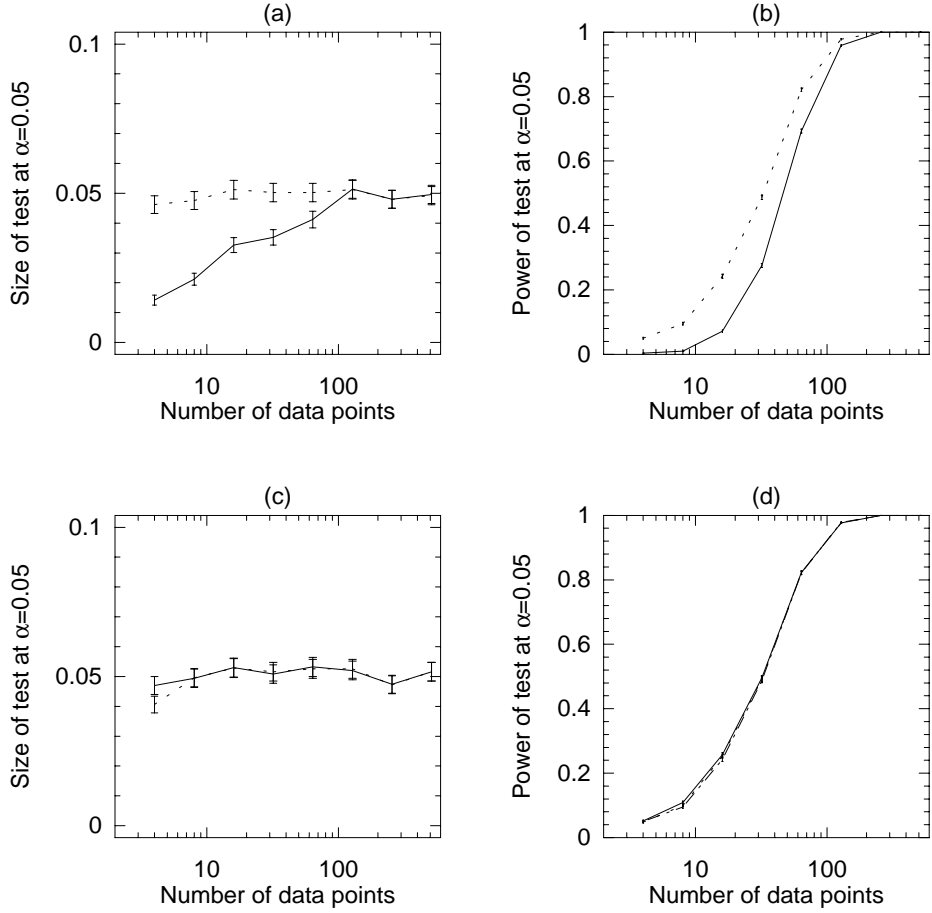

**Fig. 2.** Size and power of the test against the null hypothesis that the data is IID (independent and identically distributed) plotted against the number $n$ of points in the time series. The solid line is for the nonpivotal statistic in Eq. (8), and the dotted line is for the pivotal statistic in Eq. (9). Panels (**a,b**) are for the "typical realization" approach, and panels (**c,d**) are for the "constrained realization" approach. Plotted is the fraction of trials (out of 5000) in which the null hypothesis was rejected at the $\alpha = 0.05$ level. For the size curves, the data came from IID data generated from a normal $N(0,1)$ distribution. For the power curves, the test data came from an autocorrelated process generated by $x(i) = ax(i-1) + e(i)$ where $a = 0.4$ and the innovation $e(i)$ was IID from a normal $N(0, 1-a^2)$ distribution; note that this process has zero mean and unit variance. In the typical-realization scenario, (**a**) the pivotal statistic is more powerful than the nonpivotal statistic, and (**b**) the size of the nonpivotal statistic is smaller than the nominal value $\alpha = 0.05$ for small $n$. For the constrained-realization surrogates, both (**c**) the power and (**d**) the size of the nonpivotal statistic are very nearly the same as the pivotal statistic. Note that the pivotal statistic has the same power/size properties in the constrained-realization test as it does in the typical-realization test.

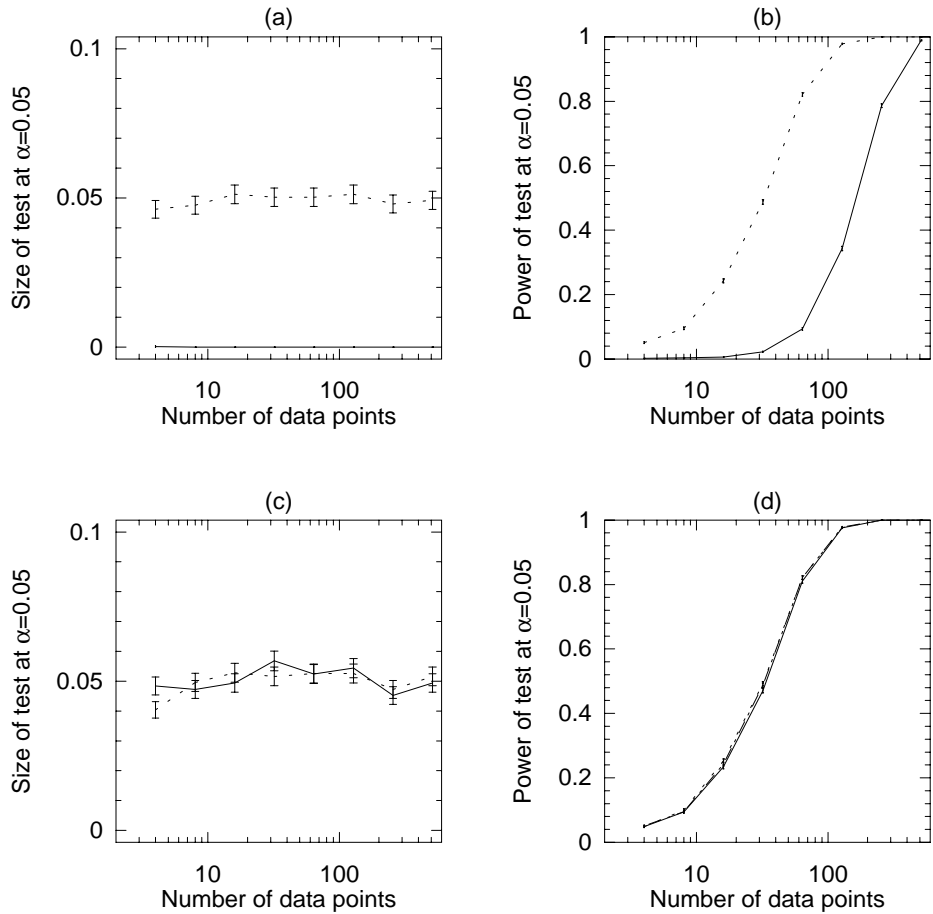

**Fig. 3.** Same as Fig. 2, except that the data for the power and size curves came from data with mean one instead of mean zero. In this case, the importance of pivotal statistics is especially dramatic, as seen in panels (**a**) and (**b**). Using the constrained-realization approach, however (**c,d**), the nonpivotal statistics perform just as well as the pivotal statistics.

identically distributed (IID). The statistic here should incorporate temporal correlations, and we consider two simple ones:

$$T = \frac{1}{n-1} \sum_{i=1}^{n-1} x(i)x(i+1). \tag{8}$$

$$T' = A(1) = \frac{1}{n-1} \sum_{i=1}^{n-1} (x(i) - \overline{x})(x(i+1) - \overline{x})/\overline{(x - \overline{x})^2}. \tag{9}$$

The second statistic is the autocorrelation at lag one; it is asymptotically pivotal in the $n \to \infty$ limit, though it is not exactly pivotal for finite $n$. Again, these are not necessarily the optimum choices for discriminating statistics, but are chosen for illustrative purposes.

The "typical realization" approach in this case is essentially Efron's bootstrap [12]. The underlying IID distribution $F$ is estimated by the empirical distribution $\hat{F}$ obtained by putting equal weight on each of the $n$ data points. A typical realization of $\hat{F}$ is achieved by resampling the data with replacement to generate a surrogate time series $X_k$.

The constrained-realization approach demands that the sample nuisance statistics agree exactly; in this case, that the sample distribution $\hat{F}$ is precisely the same for the original data as for the surrogates. Resampling *with* replacement does not achieve this, because $\hat{F}^*$ will have double weight on some points and zero weight on others where $\hat{F}$ in each case has one. However, we can make a distribution $\hat{F}'$ which matches $\hat{F}$ exactly, by resampling *without* replacement. In fact, this is nothing more than the permutation test first proposed by Fisher[24] and discussed at some length in the context of the bootstrap by Efron and Tibshirani[12]. (See also the book by Noreen [5]). This kind of temporal shuffling was also used by Schienkman and LeBaron [25] for testing IID in the context of fractal dimension estimation.

The results of two numerical experiments are shown in Fig. 2 and Fig. 3. In the first experiment, the performance of the test (as measured by accuracy and power) was estimated for a gaussian time series with mean zero and variance one. The power was estimated by testing against an alternative generated by an AR(1) process. The second experiment was the same as the first, except that both the power and size data sets had mean one instead of mean zero. For both experiments, using the typical-realization approach, the pivotal statistic out-performed the nonpivotal statistic (as it should have); the effect was especially dramatic for the second experiment because when the mean is nonzero, a pivotal (*i.e.*, mean insensitive) statistic is particularly crucial. For the constrained-realization approach, however, the nonpivotal statistic performed just as well as the pivotal statistic. This was observed in both experiments.

## 3 Testing for nonlinearity: ARMA vs FT

Serial dependence may be the *first* question to ask about a time series; a considerably less straightforward and arguably more interesing question to ask is whether it exhibits any evidence for nonlinearity. The question is motivated in no small part by the recent attention to deterministic chaos as a provocative alternative explanation for irregular (seemingly stochastic) data. Directly testing for chaos is problematic (though there is a considerable literature devoted to the problem; see Refs. [26–29] for a sampling of recent conference proceedings on the topic), but chaos requires nonlinearity, and so one may prefer first just to ask whether there is any evidence for nonlinearity. The null hypothesis, now, is that the data arise from a linear stochastic process. Two different surrogate data approaches have been identified for testing this hypothesis.[#7] One is to fit a linear model, *e.g.*, ARMA($p,q$), autoregressive moving-average:

$$x(i) = a_0 + \sum_{j=1}^{p} a_j x(i-j) + \sum_{j=0}^{q} b_j e(i-j) \qquad (10)$$

to the original data, and then by using different realizations of gaussian white noise for the residual terms ($e(i)$) one can generate an ensemble of surrogates [31–33]. It is also possible to model linear stochastic processes with a purely autoregressive (AR) or purely moving average (MA) model. The AR model is much easier to fit than either ARMA or MA models. The second is to take a Fourier transform (FT) of the data, randomize the phases, and then invert the Fourier transform to generate the surrogate data. See Ref. [34] and references therein for applications of the FT approach to potentially chaotic time series.

Refs. [35, 36] discuss some of the practical tradeoffs between FT and ARMA methods for generating surrogate data. From the point of view of this paper, the ARMA method is a "typical realizations" approach: one estimates the particular linear stochastic process, and then generates typical realizations of that estimated process. The FT method, by contrast, generates "constrained realizations," because each realization is constrained to have exactly the same sample Fourier spectrum as the original data. It bears remarking that the sample Fourier spectrum, computed directly as the magnitude of the discrete Fourier transform is in many ways a *poor* estimator of the actual underlying frequency spectrum. But this does not in itself constitute a flaw in the FT-based method of surrogate data because the actual underlying spectrum is not the ultimate goal of our calculation.

One might say that the ARMA method is appropriate for *fitting the model*; it is parsimonious, and provides a constructive description of how the data might have been generated. The FT method, by contrast, is only useful for *fitting the data*. The FT provides a terrible, atrociously overfit, model; for $n$ data points, it effectively fits $n/2$ parameters. But for the purpose of generating surrogate data to be used in hypothesis testing, we will see that this over-fitting can be a virtue.

In our previous examples (gaussian and IID), it was straightforward to convert most statistics $T$ that might be of interest into pivotal statistics. For the gaussian, this is just a matter of translating by the mean, and rescaling by the standard deviation. For the IID, as long as the statistic is expressed as a function of the *ranks* of the data values $x_i$, it will be pivotal.[#8] When testing for nonlinearity, with an eye to the alternative of chaos, one may be interested in using fairly exotic discriminating statistics, involving fractal dimensions [15, 37–41], Lyapunov exponents [42–48] or nonlinear predictors [49–52], as well as various hybrid statistics which measure determinism without directly predicting [53–57]. And it can be difficult to enforce the requirement that these discriminating statistics be pivotal. It may be easier, we argue, to make the randomization method itself "pivotal" (in a manner of speaking), by using constrained realizations and whatever discriminating statistic happens to be available or attractive.

### 3.1 Contrived example

To illustrate the point, we deliberately chose a test statistic that is highly nonpivotal, and compared the power of AR surrogates to FT surrogates. The time series was $n = 512$ points

---
[#7]See also Tong [30] and references therein for a discussion and review of more conventional tests for nonlinearity in time series.
[#8]We are grateful to Mary Thompson (personal communication) for pointing this out.

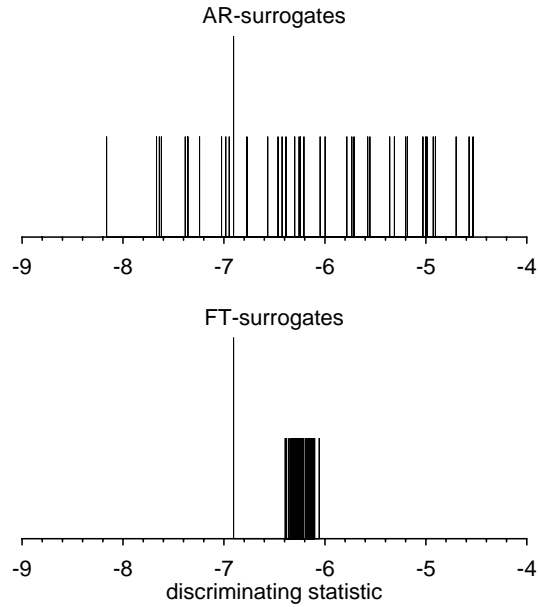

**Fig. 4.** We test for nonlinearity in a Hénon time series using the (highly contrived) discriminating statistic given in Eq. (11), and find that the test for nonlinearity based on the FT surrogates is more powerful than the same test based on AR(6) surrogates. (The order of the AR is based on the Akaike and Schwartz criteria.) In particular, the FT-based method correctly rejects the null hypothesis, while the AR method does not. The tall line is at the value of the statistic for the original data, and the short lines are at the values for the surrogate data sets.

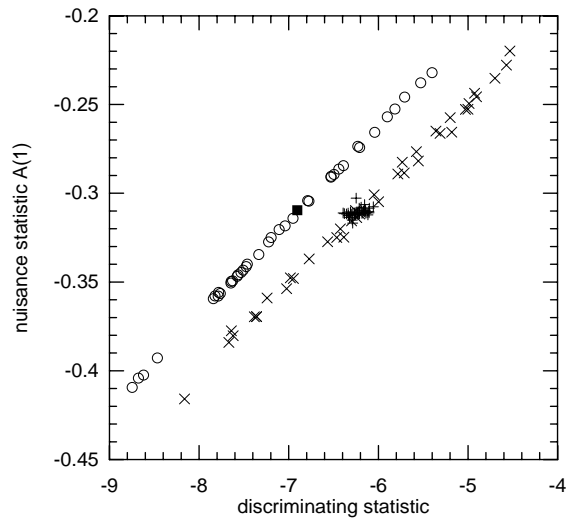

**Fig. 5.** This figure is based on the same test as in the previous figure, but attempts to clarify *why* the FT test is so much more powerful than the AR test in this case. The discriminating statistic in Eq. (11) depends very sensitively on the nuisance parameter $A(1)$ so small changes in the sample statistic for $A(1)$ generate large changes in the discriminating statistic. Here, the filled box (■) corresponds to the original data $X_o$, the open circles (○) to other realizations of the Hénon process, the crosses (×) to AR-based surrogates of the original data, and the pluses (+) to FT-based surrogates.

from the $x$-component of the chaotic Hénon map [58]. (We estimate the optimal order of the AR to be 6 based on the Akaike [59] and Schwarz [60] criteria.) As Fig. 4 shows, the FT surrogates have a much smaller variance than the AR surrogates, and are able to identify nonlinearity that AR misses. Fig. 5 shows why the AR variance is so large; the discriminating statistic in this case is

$$T = \frac{\langle (x(i+1) - x(i))^3 \rangle}{\langle (x(i+1) - x(i))^2 \rangle^{3/2}} + 20 A(1) \tag{11}$$

which is constructed to be especially sensitive to the nuisance parameter $A(1)$ given by Eq. (9). For the AR surrogates, the variance in the discriminating statistic is almost entirely due to the variance in the nuisance parameter $A(1)$ from one realization to the next. Since the FT method constrains the nuisance parameters[#9] it has considerably less variation. One may be concerned that the narrow width in the distribution of FT surrogates could lead to too many *false* rejections (or Type I errors), but we find that the FT-based test only generates false rejections about 5% of the time, which is what it should do when the significance level is set at 95%. Fig. 6 compares the size and power of this test for the two approaches.

Note that if the goal were not to test the null hypothesis but instead to get an error bar for the estimated value of some parameter, then the FT surrogates would not be appropriate. In this example, the error bar on the discriminating statistic is severely underestimated by the range of FT values, and more reasonably approximated by the values of the AR surrogates. The "actual" error bar is given by the range of values that different realizations of the Hénon process exhibit. If one is estimating a parameter which is a linear statistic, say $A(1)$, then the linear AR model does a very good job of estimating the error bar, and the FT model is worthless.

### 3.2 More realistic example

Although the statistic used in Eq. (11) illustrates what it is about the typical-realizations approach which makes it vulnerable to nonpivotal statistics, it is hardly an example that would arise in practice. We will consider another statistic which is arguably more realistic.

Since the data are suspected of being nonlinear and deterministic (or of having a deterministic component), it is natural to attempt to fit the future $x(i)$ as a nonlinear function of the past $f(x(i-1), \ldots, x(i-m))$, and to take as our statistic the average squared (in-sample) error

$$T = \frac{\overline{(x(i) - f(x(i-1), \ldots, x(i-m)))^2}}{\overline{(x - \overline{x})^2}} \tag{12}$$

For this example, we will take $m = 2$, and let $f$ be a (global) quadratic polynomial

$$f(u, v) = c_0 + c_1 u + c_2 v + c_3 u^2 + c_4 uv + c_5 v^2 \tag{13}$$

The data set itself we will also take as something a little less trivial than the simple Hénon map of the previous example; we will add five independent realizations of the Hénon map (the fractal dimension of this strange attractor is therefore five times that of a single Hénon

---

[#9] Actually, what the FT method constrains is not the ordinary autocorrelation that is plotted in the figures, but the circular autocorrelation: this treats the data as if it were periodic with period $n$, and then computes $A'(k) = (1/n) \sum_{i=1}^{n} (x(i+k \mod n) - \overline{x})(x(i) - \overline{x})/\overline{(x - \overline{x})^2}$. The difference between these different autocorrelations explains the slight variation along the vertical axis of the FT surrogates (pluses) in Fig. 5 and Fig. 8.

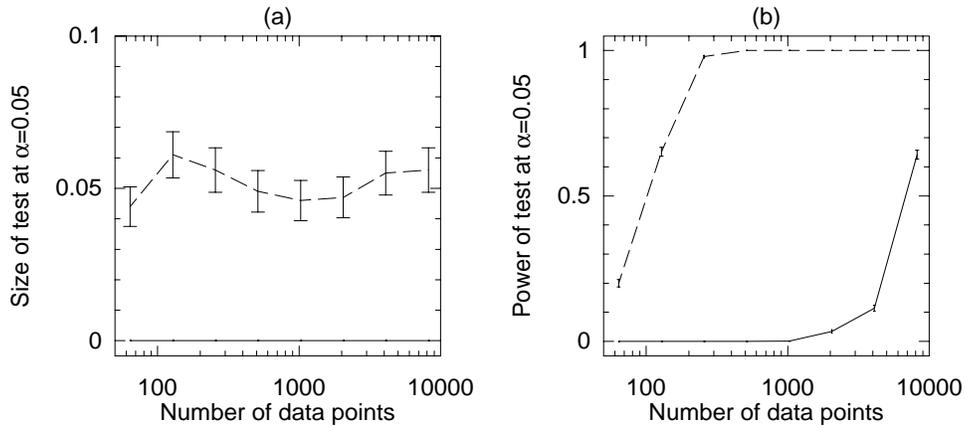

**Fig. 6.** In this figure, we compare (a) size and (b) power for Henon and AR(6) data using the "contrived" statistic in Eq. (11). Power and size are estimated as the fraction of trials (out of 1000) for which the null hypothesis is rejected at the $\alpha = 0.05$ level. The results provide a caricature of the results seen in Figs. 1,2,3; the test based on constrained-realization surrogates (dotted lines) is far more accurate and far more powerful than those based on the typical-realization surrogates (solid lines).

map, or about 5×1.2 = 6), and then we will add gaussian white noise with a standard deviation equal to 1/2 of the standard devation of the original data set. An AR(8) is used for generating the typical-realization surrogates (the order 8 is consistent with both the Akaike and the Schwartz criteria when there are $n = 512$ points in the data set), and the FT method generated the constrained-realization surrogates. As seen in Fig. 7, the AR fails to detect the nonlinearity where the FT succeeds. Again, the problem is that our "natural" choice of discriminating statisic was not pivotal; this is readily seen in Fig. 8. Like Fig. 5, the discriminating statistic is not independent of the nuisance parameters, in this case the first 2 lags of the autocorrelation, $A(1)$ and $A(2)$, and the result is a loss of power.

Accurate and powerful statistical testing for nonlinearity requires either a pivotal statistic $T$ which does not depend on $A(1)$ or any of the other autocorrelations $A(\tau)$; or a Monte-Carlo method which constrains the surrogate data to match the sample autocorrelation $A(\tau)$ of the original data. The FT-based method provides this second alternative when the statistics of interest are not pivotal.

**Acknowledgments**

We are grateful to Steve Ellner, Stephen Eubank, Andy Fraser, Danny Kaplan, Blake LeBaron, Mark Muldoon, Richard Smith, Mary Thompson, David Wolpert, and others, both for stimulating conversations on this topic and for useful comments on this manuscript. JT was partially supported by NIMH grant 1-R01-MH47184 and by the US Department of Energy. DP was supported by NSF grants ATM-9213522 and ATM-9311522.

*Finally, as an incidental but important bonus, Monte Carlo tests are relatively straightforward to explain in the course of consulting with non-statisticians.*   — Besag and Diggle, 1977
*...as long as they are not physicists.*   — Theiler and Prichard, 1993

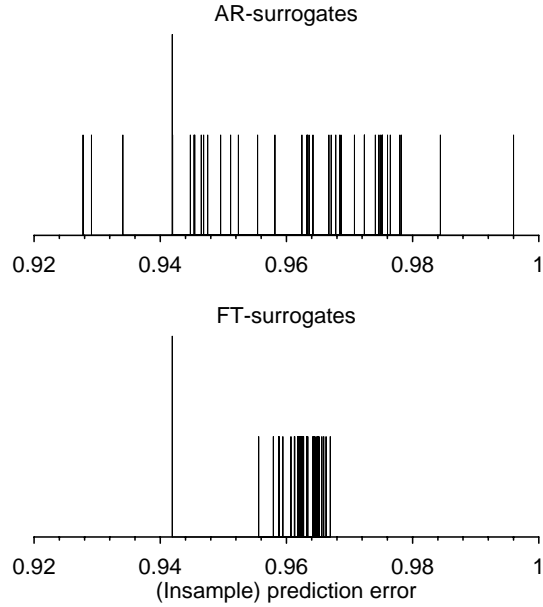

**Fig. 7.** We test for nonlinearity in a short, noisy, and relatively high-dimensional nonlinear time series. Using normalized in-sample prediction error as a discriminating statistic, we find that the FT-based surrogates detect nonlinearity, but that the AR does not.

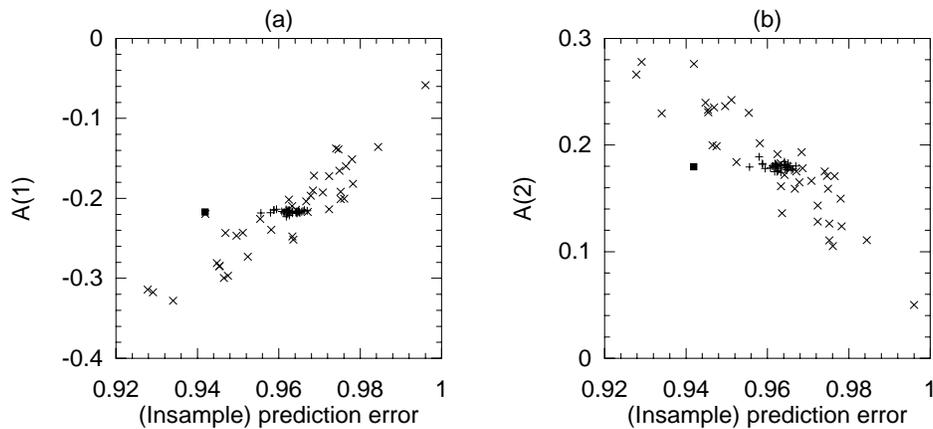

**Fig. 8.** Plotting the discriminating statistic $T$ against the nuiscance parameters $A(1)$ and $A(2)$, we can see explicitly the nonpivotal nature of this $T$. As in Fig. 5, the filled box (■) corresponds to the original data $X_o$, the crosses (×) to AR-based surrogates of the original data, and the pluses (+) to FT-based surrogates.